# Importance of Signal and Image Processing in Photoacoustic Imaging

**Rayyan Manwar** [1,¥,*], **Mohsin Zafar** [2,¥], and **Qiuyun Xu** [2]

[1]   Richard and Loan Hill Department of Bioengineering, University of Illinois at Chicago, Chicago, IL 60607, USA
[2]   Department of Biomedical Engineering, Wayne State University, Detroit, MI 48201, USA;

*   Correspondence: rmanwar@uic.edu
¥   Authors have equal contribution

**Abstract:** Photoacoustic imaging (PAI) is a powerful imaging modality that relies on the PA effect. PAI works on the principle of electromagnetic energy absorption by the exogenous contrast agents and/or endogenous molecules present in the biological tissue, consequently generating ultrasound waves. PAI combines a high optical contrast with a high acoustic spatiotemporal resolution, allowing the non-invasive visualization of absorbers at deep structures. However, due to the optical diffusion and ultrasound attenuation in heterogeneous turbid biological tissue, the quality of the PA images is deteriorated. Therefore, signal and image processing techniques are imperative in PAI to provide high quality images with detailed structural and functional information in deep tissues. Here, we review various signal and image processing techniques that have been developed/implemented in PAI. Our goal is to highlight the importance of image computing in photoacoustic imaging.

## 1. Introduction

Photoacoustic imaging (PAI) is a nonionizing and noninvasive hybrid imaging modality that has made significant progress in recent years, up to a point where clinical studies are becoming a real possibility [1-6]. Due to the hybrid nature of PAI, i.e. optical excitation and acoustic detection, this modality benefits from both rich and versatile optical contrast and high (diffraction-limited) spatial resolution associated with low-scattering nature of ultrasonic wave propagation. [7-10]. Photoacoustic imaging breaks through the diffusion limit of high-resolution optical imaging (~1 mm) by using electromagnetic energy induced ultrasonic waves as a carrier to obtain optical absorption information of tissue [11,12]. PAI, being a relatively new imaging modality, can effectively realize the structural and functional information of the biological tissue, providing a powerful imaging tool for studying the morphological structure, physiological, pathological characteristics, and metabolic functions in biological tissues [13-15].

The PA effect initiates when an optically absorbing targets (absorbers/chromophores) within the tissue are irradiated by a short (~nanosecond) pulse laser [16]. The pulse energy is absorbed by the target and converted into heat, generating a local transient temperature rise, followed by a local acoustic pressure rise through thermo-elastic expansion [17-20].

The pressure waves propagating as ultrasonic waves, are detected by ultrasonic transducers present outside the tissue, termed as raw data (Figure 1). These data carry information of inherent acoustic and optical properties (as presented in [21]) of the absorbers in combination with noisy data originated from electromagnetic interferences. The acquired data are further processed(known as signal processing) to extract the desired PA signal from the noisy background and utilized to reconstruct a PA image [22,23]. These images represent internal structures and corresponding functionality of the tissue target region [24-30]. Several image reconstruction algorithms have been studied for PA imaging [29,31-33] where the reconstruction algorithms can be interpreted as an acoustic inverse source problem [34]. Conventional PA image reconstruction algorithms assume that the object of interest possesses homogeneous acoustic properties. However, the tissue medium in reality is heterogeneous with spatially variant sound of speed and density distribution [35]. Therefore, the images produced by such algorithms contain significant distortions and artifacts [36,37]. There have been advancements of PA image reconstruction algorithms that can compensate for variations in the acoustical properties [38-43], however, further image enhancement in terms of post-processing is essential .

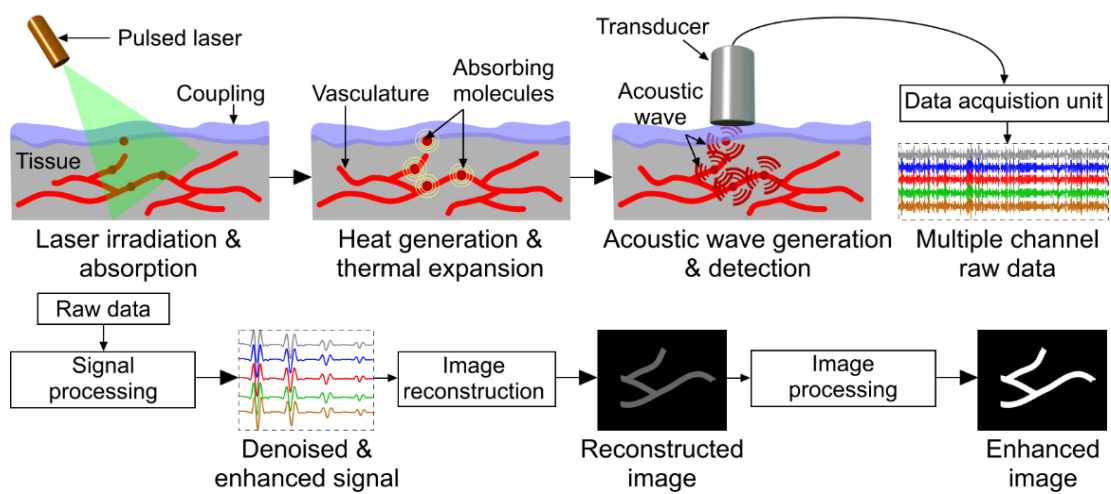

**Figure 1.** A conceptual flow of photoacoustic imaging working principle.

To obtain the morphological and functional information of the tissue chromophores, the initial goal is to retrieve the initial pressure distribution inside the object due to the absorbed laser energy. However, strong optical absorption by heterogeneous turbid superficial tissue structure pose major obstruction in irradiating the actual target located deep inside the tissue medium with sufficient optical energy [7,10,15]. Moreover, optical fluence incident upon the tissue must be limited within the pre-defined safety limit. In addition, the scattering characteristics of the tissue alters the generated PA signal [10]. Therefore, the amplitude of raw PA signal generated from deep tissue structure is very low. This limited penetration depth and optical contrast also leads to the aliasing effect. Typically, detected PA signals of ideal optical absorbing particle are of bipolar N-shape [44-46] however, the PA signals produced within a complicated biological tissue can be the combination of individual N-shape pulses from adjacent microstructures. Consequently, the PA signals from small targets are deteriorated and even buried by the bipolar signal originating from a nearby relatively large target. These phenomenon leads to aliasing and distortion in the final image [44]. In addition to the aliasing effect, the efficacy of the conversion from optical absorption to acoustic wave generation is often affected by the presence of high background noise [47,48]. The PA signals are often corrupted by background noise, from the medium and the transducer, respectively [17]. White Gaussian noise is one of the most common models for these types of randomly distributed thermal and electronic noise [49]. Furthermore, fixed-pattern noise caused by electromagnetic interference is another major source of background noise. The combination of these different types of noises offsets the PA signal, leading to a low signal-to-noise ratio (SNR), consequently producing low quality images [50-53].

Therefore, several studies have attempted to develop signal enhancement and image post-processing algorithms to either extract the original, attenuated PA signal or to improve the existing one by various filtering techniques. [50,54,55]. In many cases, these approaches were incorporated into the image reconstruction algorithms to achieve noise- and artifact-free PA images [56-62]. To further improve the prevalent image processing technique, different deep learning architectures have also been proposed [63,64].

The objective of this review article is to categorically discuss the attributes of various signal and image processing techniques used in PA. The review process is categorized into three aspects of improving PA images (i) PA signal pre-processing prior to image reconstruction and (ii) image post processing after the image reconstruction, and (iii) deep learning techniques. The search protocol used for this review study is as follows. For the first aspect, a PubMed database search of "photoacoustic" AND "signal processing" yielded 141 results with 61 published in the last five years. For the second aspect: "photoacoustic" AND "image enhancement" yielded 207 results with only 55 published in the last five years. Finally, the third aspect: photoacoustic" AND "image processing" yielded 198 articles published in the last 5 years. Among these articles 10 are associated with "image segmentation", 10 articles are relevant to "image classification". Here, we have considered only the publications where the processing concept and development methods are clearly demonstrated with appropriate experimental evaluations. Till date several major review articles have been published regarding the photoacoustic imaging and mostly are based on instrumentation and configurations. However, according to the authors knowledge, there is no dedicated review article that summarizes the key aspects of the signal and image processing in photoacoustic imaging.

Initially, we explored the root causes of diminished PA signal and corresponding degraded image quality. This follows with exploring the merit and demerits of various approaches to improve the photoacoustic signal and image quality as pre- and post-processing techniques respectively. Finally, we explore the articles where different deep learning based image processing algorithms have been utilized for improving diagnostic purpose such as classification and segmentation.

## 2. PA Signal pre-processing techniques

The complex biological tissue structures consist of several overlaying chromophores with different absorption coefficients. The PA signal from a less absorbing chromophore is either lost or overshadowed by a nearby comparatively higher absorbing chromophores. Moreover, the incident laser energy limitation imposed by ANSI (American National Standards Institute) and the optical path being highly attenuated due to the scattering in the tissue [65], results in generating a low amplitude PA signal by poorly illuminated deeper structures within the tissue. This results in PA signal being camouflaged within the background noise upon reception by the transducer, leading to a reconstruction of a very low SNR images [17]. Specifically, when a low-cost PA system based on low power LEDs are utilized, the PA signals generated from imaging target are strongly submerged in the background noise signal [35]. A typical PA signal is usually contaminated with background noises (i.e. combination of electronic and system thermal noise [48,50]. These noises are generally originated from external hardware (i.e. transducer elements, acquisition system, and laser sources). Usually, the noise from the laser source dominates at the kilohertz frequency range and attenuates following an inverse function of frequency (1/f) [66,67]. At the megahertz frequency range, , the noise from the laser source becomes less dominant [68]. Instead, the signal amplifier, the photodetector, and the data acquisition card become the major noise sources. On the other hand, biological tissue being a highly scattering medium introduce major attenuating events for the generated PA signal before it propagates and received by transducers [65]. Several pre-processing techniques to improve the PA signal to noise ratio upon reception by transducers are reviewed in the following subsections.

### 2.1 Averaging

Signal averaging is perhaps the easiest and most common way of improving the signal quality by getting rid of uncorrelated random noise. For signal averaging two schemes can be employed: (1) the raw pressure signals can be averaged coherently prior to signal processing; or (2) each of the received chirps is processed independently and the resulting correlation amplitudes are averaged [69]. These two methods define the logistics of data acquisition and may influence design of the system hardware and software for efficient signal processing. The former technique demands strict phase consistency of multiple excitation chirps and accumulation of multiple waveforms, while the latter allows for rapid processing of incoming chirps and summation of the final products to reduce noise [70]. However, the latter technique does not consider the phases of individual chirps and constitutes incoherent averaging during post-processing. Averaging specifically improves the SNR of the PA signals, particularly if the PA signal components being averaged are correlated as shown in Figure 2. A distinctive improvement by averaging method necessitates the acquisition of large number of PA signals from the same location. This acquisition number typically ranges from few hundreds to several thousands which makes this technique extremely time consuming, computationally exhaustive, and ineffective for moving targets [46].

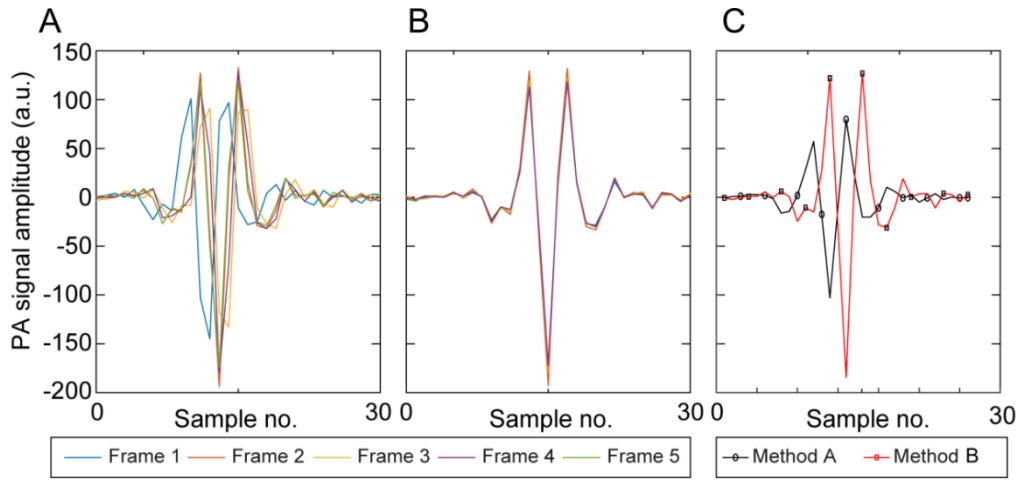

**Figure 2.** The effect of coherent PA signals on averaging technique. (A) five sequential PA signals showing lack of coherence between frames, (B) showing coherence between frames, and (C) PA signal after averaging using methods described in (A) and (B) Reproduced from [69].

### 2.2 Signal Filtering Tecnhinques:

Signal filtering techniques are often more effective when used with Fourier transformation methods. It involves selective component discarding of specific frequency bands. However, losing components of the actual PA signal along with the noise in those frequency ranges is inevitable [52]. To avoid this scenario, in [51], an adaptive and fast-filtering method to denoise and enhance the PA signal was presented. However, unlike a conventional adaptive noise canceller, this method does not require a prior knowledge about characteristics of the signal. In fact, the reference signal was basically a time shifted version of the primary input signal. Due to using a reduced number of epochs in averaging, this algorithm created a smaller PA peak time-shift and signal-broadening. A PA microscopy image with the size of $200 \times 200$ pixels using the proposed method took about 1 s, allowing near real-time PA microscopy. Najafzadeh et al. [56] proposed a signal denoising method based on a combination of low-pass filtering and sparse coding (LPFSC). In LPFSC method PA signal can be modeled as the sum of low frequency and sparse components, which allows for the reduction of noise levels using a hybrid alternating direction method of multipliers in an optimization process. Fourier and Wiener deconvolution filtering are two other common method used for PA signal denoising prior to back projection algorithm[71-73]. Typically, a window function is used to limit the signals within a specific bandwidth and leads the high-frequency components to zero [74] followed by a convolution between PA signals and illumination pulse and/ or ultrasound transducer impulse response. Wiener filter is specifically utilized to remove the additive noise. Sompel et al. [75] compared the merits of standard Fourier division technique, the Wiener deconvolution filter, and a Tikhonov L-2 norm regularized matrix inversion method. All the filters were used with the optimal setting. It was found that the Tikhonov filter were superior as compared to Wiener and Fourier filter, in terms of the balance between low and high frequency components, image resolution, contrast to noise ratio (CNR), and robustness to noise. The results were evaluated through imaging in vivo subcutaneous mouse tumor model and a perfused and excised mouse brain as shown in Figure 3A. Moradi et al. [76] proposed a deconvolution-based PA reconstruction with sparsity regularization (DPARS) technique. The DPARS algorithm is a semi-analytical reconstruction approach where the directivity effect of the transducer is taken into account. The distribution of absorbers is computed using a sparse representation of absorber coefficients obtained from the discrete cosine transform.

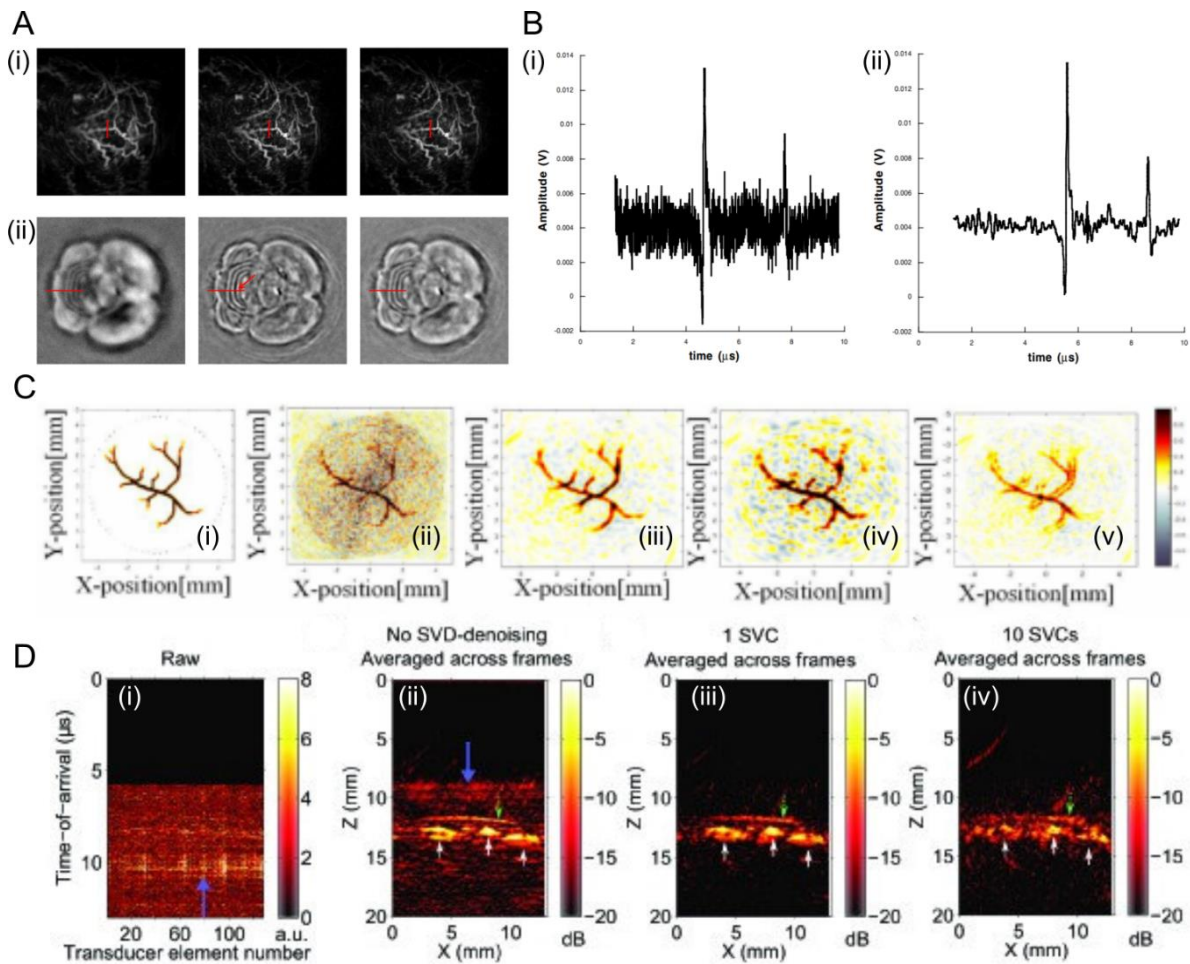

**Figure 3. (A)** Reconstructions of (i) mouse tumor and (ii) brain by deconvolution method: First column: Fourier filter. Second column: Wiener filter. Third column: Tikhonov method. The image intensities of the reconstructions are normalized (black: 0, white: 1), and the dimensions of the MIP images are 20 × 20 mm. Reproduced from [75], **(B)** (i) Photoacoustic wave generated in the blood vessel phantom prior to wavelet denoising and (ii) The processed signal using the wavelet algorithm. Reproduced from [77], **(C)** Constructed de-noising image results of PA image when simulation signal SNR is 5 dB (i) Original simulation PA image, (ii) Noisy image SNR = 5 dB, (iii) EMD combined with mutual information de-noising method, (iv) Unbiased risk estimation wavelet threshold de-noising method, (v) Band-pass filter de-noising method. Reproduced from [52] and **(D)** Laser-induced noise identification with singular value decomposition (SVD) in photoacoustic images acquired from a human finger *in vivo*, (i) in the raw radiofrequency data, vertical and horizontal noise bands were apparent [ prominent example indicated with a thick purple arrow]., (ii) When averaging across 31 PA images was performed, signals from the blood vessels were apparent but laser-induced noise across the image (prominent example indicated with a thick purple arrow) was present, (iii) When averaging across PA images and SVD-denoising with 1 SVC were performed, the laser-induced noise was absent and signals from the blood vessels and skin surface were clearly visible, and (iv) The signals from the skin surface and the blood vessels were smaller relative to the background noise when 10 SVCs were used. Reproduced from [78]

## 2.3 Transformational Techniques:

Wavelet transform based filtering techniques have become an effective denoising method. This frequency-based transform decomposes the signals into a series of basis functions with different coefficients. Usually the smaller coefficients corresponds to the noisy signals, that can be removed using thresholding [79]. In discrete wavelet transform denoising, firstly a suitable mother wavelet is selected and then decomposition, thresholding and reconstruction steps are performed. Mother wavelet selection is the most critical step and depends on the wavelet characteristics or the similarity between the signal and mother wavelet [80]. The decomposition step is carried out by selecting the appropriate degree of decomposition. In decomposition steps, low-pass and high-pass filters are used based on the characteristics of the mother wavelet. The output of these filters, respectively, are called as approximation and detail coefficients. Depending on the decomposition level, filters are applied to the detail coefficients at each step. Thresholding is a signal estimation technique and a part of the denoising step where it uses the properties of the wavelet transform [81]. Traditionally, there are soft and hard thresholding as proposed by Donoho and Johnstone [82]. In hard thresholding, the wavelet coefficients smaller than the threshold value is set to zero and higher values than the threshold stay unaltered. In the soft thresholding method, if the absolute value of the wavelet coefficients is less than or equal to the threshold value, then the coefficients are set to zero. There are different threshold selection rules (i.e. Rigrsure, Sqtwolog, Heursure, Minimaxi) [83]. Guney et al. [84] evaluated the performance of wavelet transform based signal processing methods (bior3.5, bior3.7 and sym7) in MATLAB by using the PA signals as input signals, acquired from blood vessels using photoacoustic microscopy (PAM). The results were compared with conventional FIR low and bandpass filters. Results of the LPF and BPF were very close to each other, however, sym7/sqtwolog/soft thresh. combination provided superior performance than the other two. Viator et al. [85] utilized spline wavelet transforms to enhance the PA signal acquired for port-wine stain (PWS) depth measurements. Denoising was performed in two steps: signal averaging during the experiment and post-experiment using wavelet shrinkage techniques. During the experiment, the signals were averaged over 64 laser pulses to minimize random noise. Longer averages were not taken because of dynamic processes that could change the photoacoustic signal, such as subject movement. Further denoising was accomplished with wavelet transforms using Wavelet Explorer (Wolfram Research, Inc., Urbana, Ill.), an add-on of Mathematica. Wavelet shrinkage for denoising was explained in Donoho and Johnstone [86]. Spline wavelets were chosen after verifying that the expected pressure signal was suited to relatively low order polynomial fits based on visual inspection of noisy signals. The denoising algorithm used four-level spline wavelet transforms and obtained the threshold level by estimating the noise level on each signal. The threshold was selected by taking a value between the noise level and the smallest signal variation, with the threshold set closer to the noise level (approximately 2–3 times the noise level). Holan et al. [77] proposed an automated wavelet denoising method. This approach involves using the maximal overlap discrete wavelet transform (MODWT). In contrast to the discrete wavelet transform (DWT), the MODWT yields a nonorthogonal transform. Although the MODWT requires $Nlog2N$ multiplications, versus $N$ using DWT, where $N$ is the sample size. This aspect is crucial to the extent that it eliminates one form of user intervention, such as padding with zeros or arbitrary truncation, that often occurs when using wavelet smoothing. Additionally, in contrast to the DWT, the MODWT forms a zero-phase filter making it convenient to line up features with the original signal. Here, the threshold is chosen based on the data and can be cast into a fully automatic smoothing algorithm. The benefit of this threshold is that, for large sample sizes, it guarantees that the noise will be removed with probability one. It achieved 22% improvement in the blood vessel images they reconstructed using recorded PA signals (Figure 3B). Ermilov et al. [87] implemented the wavelet transform using a wavelet family resembling the N-shaped PA signal. The wavelet transform has been established in signal processing as a superior tool for pattern recognition and temporal localization of specified signal patterns. This process helps to eliminate low-frequency acoustic artifacts and simultaneously transform the bipolar pressure pulse to the monopole pulse that is suitable for the tomographic reconstruction of the PA image. It was reported in [87] that the third derivative of the Gaussian wavelet was the best candidate for filtering the N-shaped signals. In the frequency domain, the chosen wavelet had a narrow bandpass region and a steep slope in the low-frequency band, which allowed more precise recovery of the PA signals. Based on the full understanding of PA signals features, Zhou et al. [88] proposed a new adaptive wavelet threshold de-noising (aWTD) algorithm, which provides adaptive selection of the threshold value. A simulated result showed approximately 2.5 times improvement in SNR. With wavelet denoising, signal energy is preserved as much as possible, removing only those components of the transform that exist beneath a certain threshold. This method effectively preserves signal structure, while selectively decimating small fluctuations associated with noise. Choosing the threshold is of prime importance, although an effective threshold can be chosen by simple inspection of the noisy signal [89].

*2.4 Decomposition Techniques:*

Improving the SNR of photoacoustic signal effectively is essential for improving the quality of photoacoustic image. Empirical mode decomposition (EMD) takes advantage of the time scale characteristics of data itself. It is quite suitable for non-stationary and non-linear physiological signals such as photoacoustic signals [90]. Therefore, EMD is widely used in many signal-processing fields [91-93]. In the case of noisy PA signal, EMD adaptively decomposes PA signal into several intrinsic mode functions (IMF), and remove those IMFs that is representing noise in the PA signal. Generally, if more IMFs are generated, better segregation between noisy IMFs and clean PA IMFs can be performed. An effective selection of IMFs is necessary for the successful and accurate denoising of the PA signals. Zhou et al. [52] proposed an EMD method combined with conditional mutual information denoising algorithm for PAI. Mutual information is the amount of information shared between two or more random variables. The main goal of feature selection is to use as few variables to carry as much information as possible to remove irrelevant and redundant variables. In practice, the former IMFs are mainly high frequency information and carry more noise. Therefore, it was proposed to calculate the mutual information between each of the first half of the IMF and the sum of the second half of the IMFs. When an IMF carries more unknown useful signals and less noise information, it is better to express original useful signals. According to this principle, by minimizing the mutual information between the selected IMF and the noisy PA signal, the selected mode has the most useful information. A comparative result (Figure 3C) shows that EMD combined with mutual information method improves at least 2 dB and 3 dB, respectively, more than traditional wavelet threshold method and band-pass filter. Sun et al. [94] proposed the consecutive mean square error (CMSE) based EMD method to determine demarcation point between high-frequency and low-frequency IMF. Guo et al. [95] proposed a method to improve PA image quality through signal processing method directly working on raw signals, which includes deconvolution and empirical mode decomposition (EMD). During the deconvolution procedure, the raw PA signals are de-convolved with a system dependent PSF which is measured in advance. Then, EMD is adopted to adaptively re-shape the PA signals with two constraints, positive polarity and spectrum consistence.

Another decomposition method is single value decomposition (SVD). During image reconstruction, $g=Hf$ is solved for $f$ (a finite-dimensional approximation of the unknown object(s) that produced the data in $g$) where $g$ is a vector that represents the measured data set, $H$ is the imaging operator. Ideally, $H$ would be invertible. However, it is generally found that for a real imaging system $H$ ($M \times N$ matrix) is singular. For singular matrices, it can be decomposed by means of $H = USV^T$, where $U$ is an $M \times M$ matrix, $V$ is an $N \times N$ matrix, and both are nonsingular. The $M \times N$ matrix $S$ is a diagonal matrix with non-zero diagonal entries representing the singular values of the imaging operator. The decomposition of $H$ into these component matrices is known as the singular value decomposition Each singular value of S relates the sensitivity of the imaging operator to the corresponding singular vectors in U and $V^T$. Upon decomposing the imaging operator, the vectors provided in $V^T$ are linearly independent. However, by examining the associated magnitude of the singular values in matrix $S$, it is clear not all vectors contribute equally to the overall system response. In fact, some do not effectively contribute at all to the reconstruction of an object [96]. It is the matrix rank (number of linearly independent rows) of the imaging operator that indicates the singular vectors that contribute usefully to image reconstruction. A number of techniques have been proposed to determine the rank of a matrix in the context of a real imaging operator [97-99]. Singular value decomposition (SVD) was used to identify and remove laser-induced noise in photoacoustic images acquired with a clinical ultrasound scanner [78]. The use of only one singular value component was found to be sufficient to achieve near-complete removal of laser-induced noise from reconstructed images (Figure 3D). The signals from the skin surface and the blood vessels were smaller relative to the background noise when 10 SVD components were used.

## 2.5 Other Methods

Mahmoodkalayeh et al. [65] demonstrated that the SNR improvement of the photoacoustic signal is mainly due to the reduction of Grüneisen parameter of the intermediate medium which leads to a lower level of background noise. Yin et al. [100] propose a method to optimize the speed of sound (SOS), based on a memory effect of PA signal. They revealed that the PA signals received by two adjacent transducers have a high degree of similarity in waveform, while a time delay exists between them. The time delay is related to the SOS. Based on this physical phenomenon, an iterative operation is implemented to estimate the SOS used for image reconstruction. Although PAT improved by the proposed method, artifacts and distortions still exist due to the refraction and reflection in both simulations and experiments.

## 3. Image Processing

Artifacts are one of the major problems in PAI. The presence of artifacts limits the application of PAI and creates hurdles in the clinical translation of this imaging modality. Reflection artifact is one of the most commonly observed artifacts in photoacoustic imaging [101-103]. These reflections are not considered by traditional beamformers which use a time-of-flight measurement to create images. Therefore, reflections appear as signals that are mapped to incorrect locations in the beamformed image. The acoustic environment can also additionally introduce inconsistencies, like the speed of sound, density, or attenuation variations, which makes the propagation of acoustic wave very difficult to model. The reflection artifacts can become very confusing for clinicians during diagnosis and treatment monitoring using PA imaging.

One of the standard techniques used for denoising images is wavelet thresholding. Its application ranges from noise reduction, signal and image compression up to signal recognition [104]. The advantage of this method is that the denoising approach is model-free and can be applied as a post processing step. Haq et al. [105] proposed a 3D PA image enhancement filter based on Gabor wavelet integrated with traditional hessian filter to clearly visualize the vessels inside mouse brain with scalp open. In the proposed method Gabor wavelet filter is used to enhance the vasculature (Figure 4), then hessian-based method is applied to classify vessel-like structures in the PAM generated image.

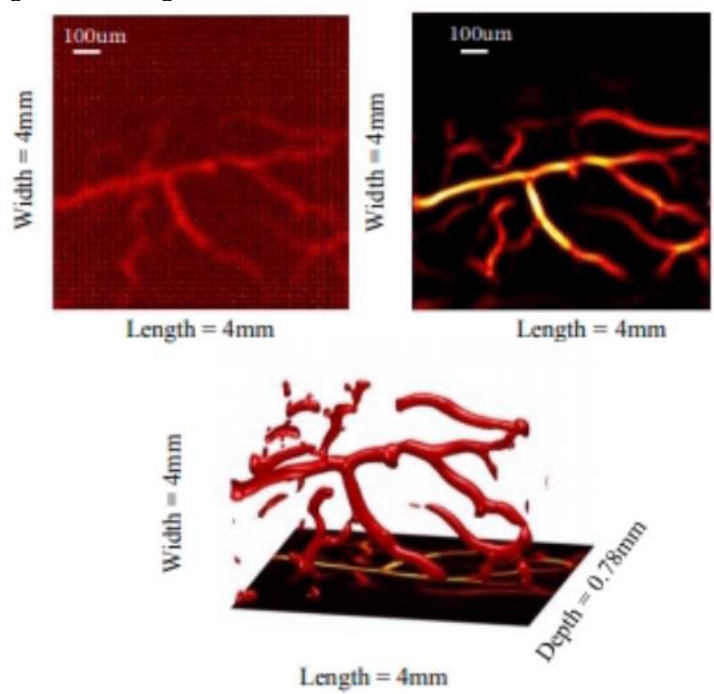

**Figure 4.** Original image of living mouse brain vessels (top left); Filtered MIP of photoacoustic image of vessels (top right), and 3D reconstruction of the vasculature (bottom). Reproduced from [105].

Deconvolution algorithms have proved instrumental in improving the quality of PA images as well. Many studies showed the effectiveness of the deconvolution-based PA image reconstruction[72,73,106-109]. Deconvolution algorithm has been used to remove the artifacts caused by the pulse width of the laser and bandwidth of the transducer [110]. Deconvolution algorithms are also used for deblurring purposes [111]. The blurry artifacts are very common in PA images and usually introduced by the inherent characteristics of the optical setup. These artifacts are due to the spatial non-uniformity of the laser beam size, poor or unoptimized optical alignment or low-quality lenses. To get rid of the blurring artifacts, a very fine structure is imaged and point spread function (PSF) is computed. The acquired images from the system are deconvolved with PSF to remove deblurring artifacts. However, PSF only provides blurring and aberration information based on the optics of the system. Since PAI is a hybrid technique, blurring and aberration caused by acoustic focus must also be considered. Seeger et al [112] introduced high-quality total impulse response (TIR) determination based on spatially-distributed optoacoustic point sources (SOAPs). The SOAPs are produced by scanning an optical focus on an axially-translatable 250 nm gold layer. This TIR method includes the optical impulse response describing the characteristics of optical excitation, the spatial impulse response (SIR) capturing the spatially-dependent signal modification by the ultrasound detection, and the spatially-invariant electric impulse response (EIR) embodying the signal digitization [113-115]. Using a spatially dependent TIR-correction improved the SNR by >10 dB and the axial resolution by ~30%. A comparison between conventional reconstruction and TIR correction was performed for an isolated RBC in vitro (Figure 5A), which was imaged at the acoustic focus. Wang et al. [116] also showed that PAI spatial resolution can be enhanced with impulse responses. However, in contrast to the SIR, finding the EIR is challenging [117].

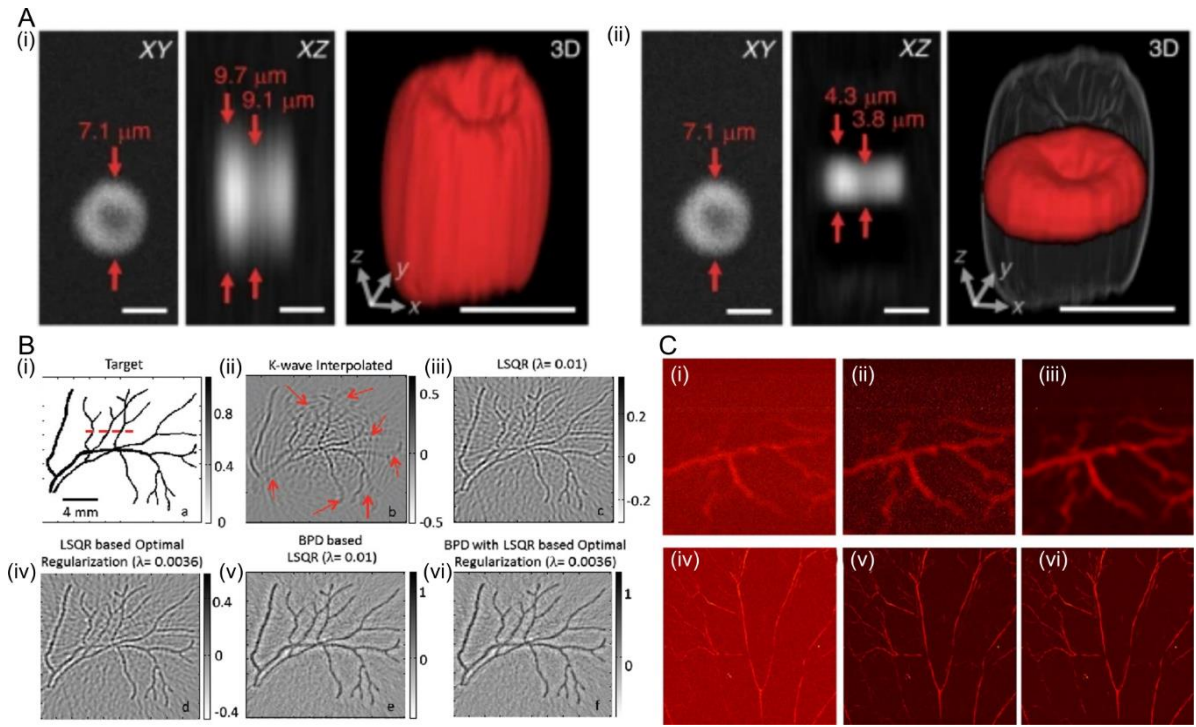

**Figure 5. (A)** Conventional reconstruction leads to an axially elongated RBC; f the TIR-corrected RBC appears flatter and smoother. Reproduced from [112], **(B)** Reconstructed photoacoustic images of (i) the target using (ii) k-wave interpolated, (iii) LSQR with heuristic choice of λ, (iv) LSQR with optimal choice of λ, (v) Basis pursuit deconvolution (BPD) with heuristic choice of λ in LSQR framework, and (vi) BPD with optimal choice of λ in LSQR framework. Reproduced from [111], and **(C)** (i) Raw images, (ii) band pass filtered images, and (iii) NLMD images. Top row: In vivo mice brain images, bottom row. Reproduced from [118].

In [111], basis pursuit deconvolution (BPD) was utilized to deblur the solution obtained using the Lanczos-Tikhonov regularization method. As regularization blurs the solution, the effect of regularization can be overcome by the BPD method. BPD utilizes the split augmented Lagrangian shrinkage algorithm (SALSA) [119] to minimize the objective function, which uses ℓ1-type regularization to promote sharp features. A numerical blood vessel phantom as shown in Figure 5B with initial pressure rise as 1 kPa was also used to demonstrate the performance of the algorithm. It was also shown that using the proposed framework, the quantitative accuracy of the reconstructed photoacoustic image has improved by more than 50%. Lucy-Richardson iterative deconvolution algorithm is another common method for removing PA blurring artifacts. Cai et al. [120] developed an iterative algorithm based on the Lucy-Richardson (LR) deconvolution with a known system PSF. The iterative equation to seek the optimal estimation of original image was derived from the maximum-likelihood estimate approach. The lateral and axial resolution was improved by 1.8 and 3.7 times and the axial resolution by 1.7 and 2.7 times that was evaluated by imaging *in vivo* imaging of the microvasculature of a chick embryo.

The other standard method for denoising images is non-local means (NLM) filtering [118,121]. Like wavelet denoising methods, this also does not rely on any imaging model and can be applied as a post-processing method that are corrupted with Gaussian noise. The principle of NLM denoising is taking the average intensity of the nearby pixel weighted by their similarity [121-123]. In [118], the objective was to remove noise from PA images and estimating the effective proposed denoising input parameters. Authors have shown that the noise was reduced and the contrasts between vessel and background were higher when NLM process was utilized as compared to the band pass filtered images.as shown in Figure 5C.

Awasthi et al. [124] proposed a guided filtering approach, which requires an input and guiding image. This approach act as a post processing step to improve commonly used Tikhonov or total variational regularization method. The same guided filtering [125] based approach has been used to improve the reconstruction results obtained from various reconstruction schemes that are typically used in PA image reconstruction.

## 3. Deep Learning for Image Processing

Deep learning (DL) approach is also used for photoacoustic imaging from sparse data. In DL, linear reconstruction algorithm is first applied to the sparsely sampled data and the results are further applied to a convolutional neural network (CNN) with weights adjusted based on the training data set. Evaluation of the neural networks is non-iterative process and it takes similar numerical effort as a traditional back projection algorithm for photoacoustic imaging. This approach consists of two steps: In the first step, a linear image reconstruction algorithm is applied to the photoacoustic images, this method provides an approximate result of the original sample including under-sampling artifacts. In the next step, a deep CNN is applied for mapping the intermediate reconstruction to form an artifact-free image [59]. Antholzer et al. [59] demonstrated that appropriately trained CNNs can significantly reduce under sampling artefacts and increase reconstruction quality (Figure 6A).

Zhang et al. [126] implemented a pre-processing algorithm to enhance the quality and uniformity of input breast cancer images and a transfer learning method to achieve better classification performance. The traditional supervised learning method was initially applied to photoacoustic images of breast cancer generated in K-wave simulation, extracted the Scale-Invariant Feature Transform (SIFT) features, and then used K-means clustering to obtain the feature dictionary. The histogram of the feature dictionary was used as the final features of the image. Support Vector Machine (SVM) was used to classify the final features, achieving an accuracy of 82.14%. In the deep learning methods, AlexNet and GoogLeNet are used to perform the transfer learning, achieving 89.23% and 91.18% accuracy, respectively. Finally, the authors concluded that the combination of deep learning and photoacoustic imaging can achieve higher diagnostic accuracy than traditional machine learning based on the comparison of the Area Under the Curve (AUC), sensitivity, and specificity among SVM, VGG, and GoogLeNet [127-130]. Chen et al developed a deep-learning-based method to correct motion artifacts in optical resolution photoacoustic microscopy (OR-PAM). The method established an end-to-end map from input raw data with motion artifacts to output corrected images. Vertical, horizontal, and complex pattern motion artifacts were introduced on PAM images of rat brain. The images with the motion artifacts were used for training and original images were considered as ground truth. The trained neural network was able to remove motion artifacts in all direction [131].

Use of directly reconstructed images on the neural networks to remove artifacts is a valid approach in many applications, specifically if the goal is to achieve fast and real-time reconstructions. This approach only needs an initial direct reconstruction and a trained network. In the case of a full-view data, this is a promising approach, but it has been demonstrated that even with limited-view images this technique performs very well to enhance the image quality [132].

The densenet-CNN accepts a low-quality PA image as input and as output generates high quality PA image [61]. One of the major advantages of using the dense convolutional layer is that it utilizes all the generated features from previous layers as inputs through skip connections. This enables the propagation of features more effectively through the network which leads to the elimination of the vanishing gradient problem. To obtain the output image, all the features from the dense blocks are concatenated, a single convolution with one feature map is performed at the end. Sushanth et at. [133] used dictionary-based learning (DL) methods to remove reverberation artifacts that obscure underlying microvasculature. Briefly, signals obtained at depths in PAM systems are often obscured by acoustic reverberant artifacts from superficial cortical layers, therefore, cannot be used. The developed DL method demonstrated suppressing of reverberant artifacts by $21.0 \pm 5.4$ dB, enabling depth-resolved PAM up to 500 μm from the brain surface of a live mouse.

Manwar et al. [66] trained a U-Net with a perceptually sensitive loss function to learn how to enhance the low signal-to-noise ratio (SNR) structures in a PA image that are acquired with a low energy laser where the high energy images used as label. After the enhancement, outline of the deeper structures such as lateral ventricle, third ventricle became more prominent in in-vivo sheep brain imaging. Hariri et al. [134] proposed a denoising method using a multi-level wavelet-convolutional neural network (MWCNN) to map low fluence illumination source images to its corresponding high fluence excitation map. In this setting, the model was inclined to distinguish noise from signal based on the shape features. The model was trained in a supervised manner to transform low energy inputs into outputs as close as possible to the ground truth frames. Substantial improvements up to 2.20, 2.25, and 4.3-fold for PSNR, SSIM, and CNR metrics were observed. In an *in vivo* application the proposed method enhanced the contrast up to 1.76-times. Reconstructed images and corresponding CNR is shown in Figure 6C. Rajanna et al [135] proposed a combination of an adaptive greedy forward with backward removal features selector along with a deep neural network (DNN) classification.

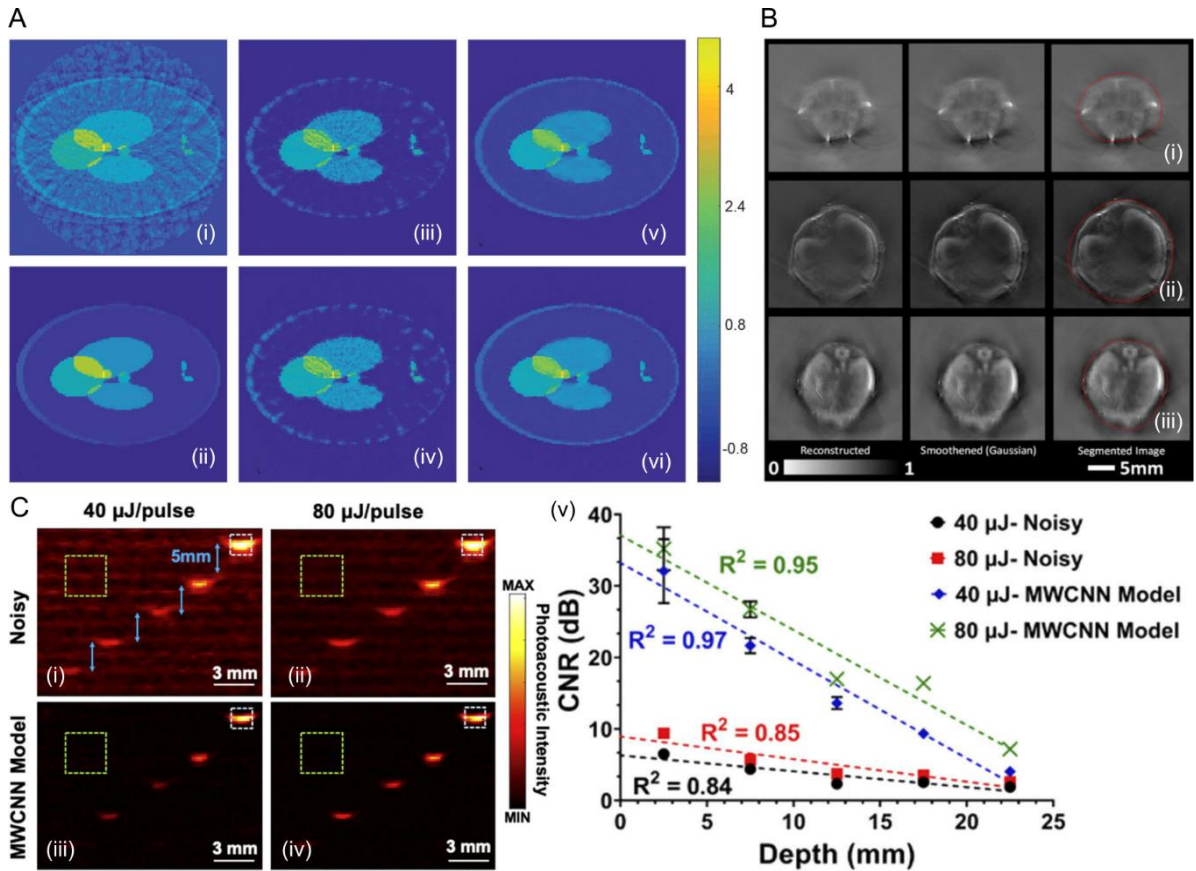

**Figure 6.** **(A)** Reconstruction results for a Shepp–Logan type phantom from data with 2% Gaussian noise added. (i) FBP reconstruction; (ii) reconstruction using TV minimization; (iii) proposed CNN using wrong training data without noise added; (iv) proposed CNN using wrong training data with noise added; (v) proposed CNN using appropriate training data without noise added; (vi) proposed CNN using appropriate training data with noise added. Reprinted from [59], **(B)** Tomographic optoacoustic reconstructions of the brain (i), liver (ii) and kidney/spleen (iii) regions of mice in vivo. The original reconstructed images obtained with model-based inversion are shown in the first column. The second column displays the smoothened images after Gaussian filtering. The segmented images using active contour (snakes) with the optimum parameters are showcased in the third column. Reproduced from [136], and **(C)** (i) B-mode noisy (input) photoacoustic image using LED at a fluence of 40 μJ/pulse. Pencil leads were placed at 2.5, 7.5, 12.5, 17.5, and 22.5 mm in 2% intralipid. (ii) B-mode noisy (input) photoacoustic images at a fluence of 80 μJ/pulse with similar experimental setup as described in (ii). (iii) and (iv) B-mode MWCNN model (output) photoacoustic image for 40 and 80 μJ/pulse. (v) CNR versus depth for 40 and 80 μJ/pulse in both noisy and MWCNN model. Dotted green and white rectangles represent the ROI used to measure mean values and standard deviations of background. Reproduced from [134].

Image segmentation is often challenged by low contrast, noise, and other experimental factors. In pulse-echo US images, common artifacts are related to attenuation, speckle noise or shadowing, which may result in missing boundaries [137]. Efficient segmentation of multi spectral optical tomography images is similarly hampered by the relatively low intrinsic contrast of large anatomical structures and tissue boundaries [138]. Mandal et al. [136] and Lafci et al. [139] used an active contour edge detection algorithm received as input PA or US images as a square array of 256x256 pixels. The images were first downscaled to 150x150 pixels to reduce the computation time whereas the pixel intensities were converted to 8-bit range between 0 and 255. Edge detection was implemented to overcome any dependency of the initial guess upon the user. Canny edge detector [36] was applied after smoothing the image using Gaussian filter with kernel size 3 and sigma 0.5. The outliers and the non-connected components in the pixels erroneously detected as edges were removed by applying morphological operations of dilation and erosion with a disc-shaped structuring element of 3-pixel size. Specifically, the segmented boundary information was used to aid automated fitting of the SOS values in the imaged sample and the surrounding water. A reconstruction mask was further used for quantified mapping of the optical absorption coefficient by means of light fluence normalization. The performance of active contour segmentation for cross-sectional optoacoustic images and the associated benefits in image reconstruction were demonstrated in phantom and small animal imaging experiments.

## 4. Conclusions

PA imaging is an emerging non-invasive hybrid modality with advantage of optical contrast and acoustic spatial resolution. Despite the advantages, PA imaging needs more refinements before its clinical translation. One of the primary issues with PA imaging is that its efficiency is limited by the presence of background noise and that PA signals suffer from low SNR which subsequently leads to degraded image quality. Therefore, utilization of PA signal processing as well as image enhancement algorithms to improve the quality of PA imaging are essential. Here, we discussed major signal processing techniques used in PA imaging, including conventional and adaptive averaging, signal deconvolution, wavelet transform, single value decomposition, and empirical mode decomposition. The signal processing techniques have been utilized to primarily denoise the PA signal before feeding them into a reconstruction algorithm. Existing reconstruction algorithms have their own merits and demerits. However, in most cases, due to inherent limited view problem and partial considerations of actual acoustic medium, the reconstruction methods are unable to represent the features of the imaging target, accurately. There have also been several studies investigating PA image post-processing such as enhancement, segmentation, classification for the purposes of disease detection of staging of the disease. Some of these algorithms are such as: wavelet thresholding, active contour segmentation, basis pursuit deconvolution, non-local mean algorithms. In addition to conventional data or image curation techniques, deep learning based signal and image processing have recently gained much popularity, specifically for obtaining high quality PA image. These techniques were also discussed in details. This study showed that PA signal processing has certainly improved the SNR of the signal in larger depths similar to when higher energy laser is used. It also showed that image post-processing algorithms improve the diagnostic capability of PA imaging.

**Conflicts of Interest:** The authors declare no conflict of interest